\documentclass[preprint,review,11pt]{elsarticle}
\usepackage{graphicx}
\usepackage{amssymb}
\newcommand{\Msun}{\ensuremath{M_{\odot}}}

\newcommand{\Ha}{H$\alpha$ }

\newcommand{\arcsec}{\ensuremath{^{\prime\prime}}}
\newcommand{\arcmin}{\ensuremath{^{\prime}}}
\newcommand{\mnras}{MNRAS}

\journal{New Astronomy}
\begin{document}
\begin{frontmatter}
\title{Dust extinction and X-ray emission from the star burst galaxy NGC 1482}
\author[]{N. D. Vagshette$^1$}
\author[]{M. B. Pandge$^1$}
\author[]{S. K. Pandey$^2$}
\author{M.K.~Patil$^1$\corref{cor1}}
\ead{patil@iucaa.ernet.in}
\address{$^1$School of Physical Sciences, S.R.T.M. University, Nanded-431 606 (MS), India Tel.+91-2462-229242; +91-8308298063/Fax:+91-2462-229245}
\address{$^2$School of Studies in Physics \& Astrophysics, Pt. Ravishankar Shukla University, Raipur-492 010 (CG), India}
\cortext[cor1]{Corresponding author}

\begin{abstract}
We present the results based on multiwavelength imaging observations of the prominent dust lane starburst galaxy NGC 1482 aimed to investigate the extinction properties of dust existing in the extreme environment. (B-V) colour-index map derived for the starburst galaxy NGC 1482 confirms two prominent dust lanes running along its optical major axis and are found to extend up to $\sim$ 11\,kpc. In addition to the main lanes, several filamentary structures of dust originating from the central starburst are also evident. Though, the dust is surrounded by exotic environment, the average extinction curve derived for this target galaxy is compatible with the Galactic curve, with $R_V$=3.05, and imply that the dust grains responsible for the optical extinction in the target galaxy are not really different than the canonical grains in the Milky Way. Our estimate of total dust content of NGC 1482 assuming screening effect of dust is $\sim 2.7\times10^5$ \Msun, and provide lower limit due to the fact that our method is not sensitive to the intermix component of dust. Comparison of the observed dust in the galaxy with that supplied by the SNe to the ISM, imply that this supply is not sufficient to account for the observed dust and hence point towards the origin of dust in this galaxy through a merger like event. 

Our multiband imaging analysis reveals a qualitative physical correspondence between the morphologies of the dust and \Ha emission lines as well as diffuse X-ray emission in this galaxy. Spatially resolved spectral analysis of the hot gas along outflows exhibit a gradient in the temperature. Similar gradient was also noticed in the measured values of metallicity, indicating that the gas in the halo is not yet enriched. High resolution, 2-8\,keV \textit{Chandra} image reveals a pair of point sources in the nuclear region with their luminosities equal to 2.27 $\times$10$^{39}$erg s$^{-1}$ and 9.34$\times10^{39}$ erg s$^{-1}$, and are in excess of the Eddington-limit of 1.5\Msun accreting source. Spectral analysis of these sources exhibit an absorbed-power law with the the hydrogen column density higher than that derived from the optical measurements.
\end{abstract}


\end{frontmatter}


\section{Introduction}
Dust is a crucial component of the interstellar matter (ISM). It provides site for interstellar chemistry and also regulates the thermal balance in various phases of the ISM. It also provides shielding for the dense clouds and hence critically influence the star formation processes. The characteristics of interstellar dust strongly depend on the balance between its formation and destruction, and on the processing of dust grains in the ISM. In the starburst galaxies, where active star formation is taking place, dust grains are not only produced and ejected from stars but are also processed in the ISM. During their starburst phase, the supernova rate (SNe) are enhanced, and hence the hot gas swept up by the reverse and forward shocks propagating within SNe can efficiently destroy the dust grains of larger size ($> 0.1\,\mu$m) (\citep{1996ApJ...469..740J}, \citep{2010PhRvD..81h3007N}). Thus, the fate of dust grains in the starburst galaxies is highly uncertain. 

Extinction curves, wavelength dependent nature of scattering and absorption of stellar light by dust grains, in optical/UV regime are the prime diagnostic tool to investigate dust properties in the extragalactic environment. They sensitively depend on the composition of dust grains and hence will allow us to estimate the nature as well as origin of dust grains in the external galaxies. Furthermore, to constrain the star formation activities, it is important to accurately determine the amount of dust as well as wavelength dependent nature of dust extinction in this class of galaxies. Though, interstellar dust is an important component, the effective dust extinction of the incident radiation is one of the least understood phenomenon in the extragalactic environment. The difficulties involved in understanding the dust extinction are due to the lack of the knowledge of (i) physical properties of the grains and (ii) their spatial distribution in the external galaxies (\citep{2000ApJ...533..682C}). Given the spatial distribution of dust, mapped through high resolution multiband data, the foreground dust screen model proposed by \citep{2001PASP..113.1449C} can be used to derive the extinction law. 

\begin{table}
\flushleft
\caption{Global parameters of NGC 1482}
\begin{tabular}{@{}ll@{}}
\hline
\hline
Parameter & Value\\
\hline
Alternate names&	ESO 549-33; IRAS03524-2038;\\
&	MCG-3-10-54; PGC 14084\\
RA \& DEC (J2000.0) &03:54:38.9; -20:30:09\\
Morph Type & SA0+ pec\\
Mag($B_T$) & 13.10\\
$M_B$& -22.2\\
Size& 2\arcmin.54$\times$1\arcmin.4\\
Distance(Mpc)& 27.37 ($H_0$ = 70 km s$^{-1}$Mpc$^{-1}$) \\
Redshift(z) & 0.0064\\
Radial Velocity(km s$^{-1}$) & 1916$\pm$39\\ 
IRAS flux density (Jy)	& 1.55$\pm$0.02 (12$\mu$m); 4.73$\pm0.05$ (25$\mu$m);\\
&	35.33$\pm$0.06 (60$\mu$m); 45.32$\pm$0.05 (100$\mu$m)\\
SEST flux density ($m$Jy) &	143$\pm$15.7 (1300$\mu$m)\\
\hline
\end{tabular}
\label{para}
\end{table}

Starburst galaxies while passing through the intense star formation in the central region show enhanced supernova rate. Ejecta from these supernovae combine with the winds and the material ejected from the massive stars in the surrounding area and result in to the expanding super-bubbles around the super star clusters \citep{2004MNRAS.348..406H}. Present day observing facilities in the X-ray domain i.e., \textit{Chandra}, XMM-\textit{Newton} and \textit{Suzaku}, have provided the best mean to study the spatial distribution, temperature and metallicity of hot gas along these outflows (\citep{2004ApJS..151..193S}, \citep{2004ApJ...606..829S}, \citep{2004MNRAS.354..259R}, \citep{2007PASJ...59S.269T}, \citep{2008MNRAS.386.1464R}). X-ray observations of starburst galaxies have also confirmed extended halo of soft X-ray around majority of the starburst galaxies (e.g., \citep{2004ApJ...606..829S} and references therein).  In edge-on starburst galaxies, diffuse X-ray emission is traced many kiloparsecs out into the galactic halo and is believed to be the manifestation of the ``feedback'' from starbursts. Short-lived massive stars in the starburst inject kinetic energy as well as metal enriched gas into their surroundings through the stellar winds and supernovae in the form of ``superwind’’ \citep{1990ApJS...74..833H}.
\begin{figure}
\center
\includegraphics[width=80mm,height=60mm]{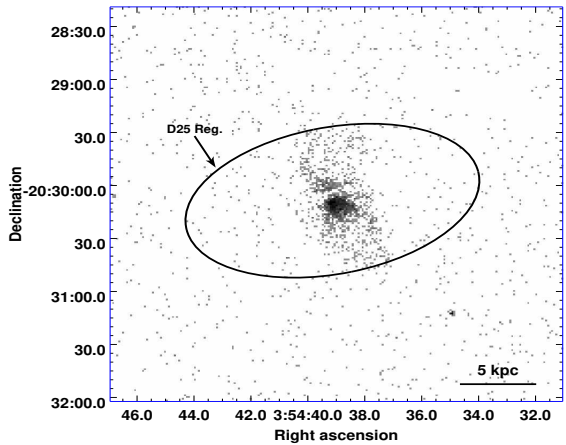}
\caption{0.3-8\,keV raw \textit{Chandra} ACIS-S3 image of NGC 1482 at full resolution of 0.\arcsec49/pixel. X-ray absorption due to dust grains along optical major axis is evident in this figure}
\label{fig1}
\end{figure}

In this paper we report on the extinction properties of interstellar dust in the starburst galaxy NGC 1482 and also on the association of dust with other phases of ISM. This galaxy has been detected in the $IRAS$ Survey, with 60 $\mu$m and 100 $\mu$m flux densities equal to 35.33$\pm$0.06 Jy and 45.32$\pm$0.05 Jy, respectively. The larger value of $S_{60 \mu m }/S_{100 \mu m}\sim 0.77$ are indicative of its starburst activity \citep{2004ApJS..151..193S}. This galaxy is defined as IR warm due to the higher value of dust grain temperature ($T_d = 44$ K) \citep{1989ApJS...70..699Y} and is relatively larger than observed in normal galaxies. This is IR bright galaxy with log(L$_{IR}$/L$\odot$) = 10.5 \citep{1989AJ.....98..766S}, rich in molecular gas and dust \citep{1989ApJS...70..699Y} and shows signatures of vigorous star formation \citep{2000ApJ...539..641T}. On the basis of the prominent dust lane oriented along its optical major axis, this galaxy is classified as peculiar galaxy in the \textit{Third Reference Catalogue of Bright Galaxies} \citep{1991Sci...254..592D}. In the emission line survey of early-type spiral galaxies \citep{1999AJ....118..730H}, have investigated the presence of filaments \& chimneys of ionised gas extending along minor axis of this galaxy \citep{1999AJ....118..730H}. On the basis of emission line imaging analysis \citep{2002ApJ...565L..63V} have demonstrated the presence of remarkable hour glass-shaped optical emission line outflow from nuclear region of this galaxy. The soft X-ray image presented by \citep{2004ApJ...606..829S};\citep {2004ApJS..151..193S} also exhibits bipolar emission along minor axis associating with the optical hour glass-shaped structure. Using the multi-frequency radio observations \citep{2005MNRAS.356..998H} have reported detection of continuum emission from the central region of this galaxy. Thus, NGC 1482 hosts multiphase ISM and is a potential candidate for the multiwavelength study. Global parameters of NGC 1482 are given in Table~\ref{para}.

The paper is structured as follows. In Section 2 we describe the multiband imaging observations and its reduction process. In section 3 we describe the morphology, extinction properties of dust and its content in NGC 1482. In this section we also describe the X-ray emission properties of NGC 1482. Section 4 describes our results derived from the present work and multiphase association of ISM in the target galaxy. Finally we summarise our results in Section 5.
\section[]{Observations and data reduction}
\subsection{Optical data}
Deep CCD images on the program galaxy NGC 1482 were obtained  using the 2.0\,m Optical Telescope of IUCAA Girawali Observatory (IGO), Pune, India,  during February 2010. These observations were performed through the broad-band filters B, V, R, I as well as narrow band filter centred on the \Ha emission using the EEV 2k$\times$2k thinned, back illuminated IUCAA Faint Object Spectrograph CCD Camera (IFOSC). This CCD provides an effective FOV of $\sim10.\arcmin5\times 10.\arcmin5$ on the sky corresponding to a plate scale of 0.\arcsec3/pixel. The seeing was 1\arcsec.5 in B band and 1\arcsec.2 in V band and the observations were performed under photometric conditions. Exposure times in different pass bands were 3000\,s\,(B), 3000\,s\,(V), 2400\,s\,(R), 1800\,s\,(I) and 4500\,s\,(H$_\alpha$) and were usually split into 3 to 5 separate exposures. In addition to the science frames, for calibration purpose, several twilight sky flats in each filter and zero exposure bias frames were also taken during this observing run. 

These observations were reduced using standard tasks, such as \textit{bias subtraction, flat fielding}, etc. available within the Image Reduction and Analysis Facility (IRAF) software. Multiple frames taken in each passband were geometrically aligned to an accuracy better than one tenth of a pixel using the centroids of the common stars in the science frames and were then co-added to improve the signal-to-noise ratio. Sky background estimation was carried out using the box method \citep{1998A&A...333..803S} and was then subtracted from the respective passband image.

Reduction of narrow band images centred on \Ha emission from NGC 1482 were done in the usual manner as discussed in \citep{2009arXiv0901.1747P} and using the tasks available within IRAF.
\subsection{X-ray data}
With an objective of studying association of dust and ionised gas with the hot, X-ray emitting gas in this galaxy, we have used the high resolution X-ray data on NGC 1482 available in the archive of {\it Chandra} X-ray observatory. NGC 1482 was observed by \textit{Chandra} (Obs. ID 2923) with the ACIS-S3 detector on the {\emph Chandra} X-ray Observatory for an effective exposure time of 28.56\,ks. Figure~\ref{fig1} shows the broadband (0.3-10\,keV) \textit{Chandra} raw image of NGC 1482. From this figure it is clear that the X-ray structure of NGC 1482 is rather complex with a bright nuclear region and extended diffuse X-ray emission. Absorption of X-ray photons along the dust lane is also evident in this figure. The data were reprocessed using the standard tasks available within the \textit{Chandra} Interactive Analysis of Observations\footnote{http://cxc.harvard.edu/ciao} (CIAO version 4.2.0) and CALDB (version 4.3.0) provided by the  \textit{Chandra} X-ray Centre (CXC). Periods of high background data were filtered out using 3$\sigma$ clipping of the 0.3-10.0\.keV light curve extracted from this chip in 200\,s bins. After removing these periods of high background, the cleaned data have a net exposure time of 24.0\,ks. The data were then filtered to contain counts in the energy range 0.3-8.0\,keV and were also background subtracted using the appropriately scaled $CXC$ background event data set. After generating an exposure map, the point sources present in the data were detected using the \textit{wavdetect} tool available within the CIAO adopting a threshold significance of 1.8$\times 10^{-6}$. A total of 12 point sources were detected on the S3 chip, however, except the two nuclear sources no other point source was found within the optical D$_{25}$ ellipse \citep{1991Sci...254..592D}.

\begin{figure}
\center
\includegraphics[width=40mm,height=39mm]{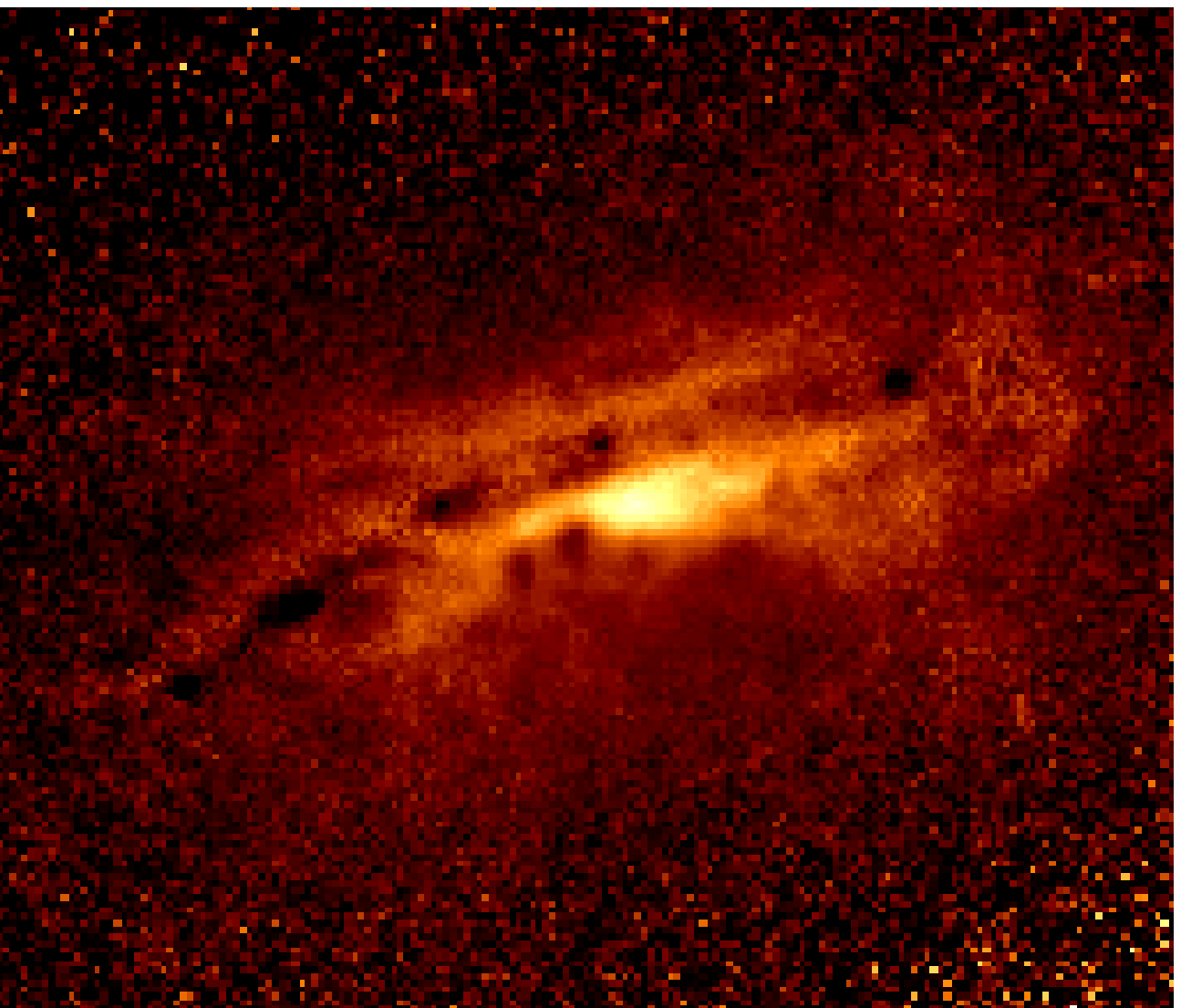}
\hspace{5mm}
\includegraphics[width=40mm,height=40mm]{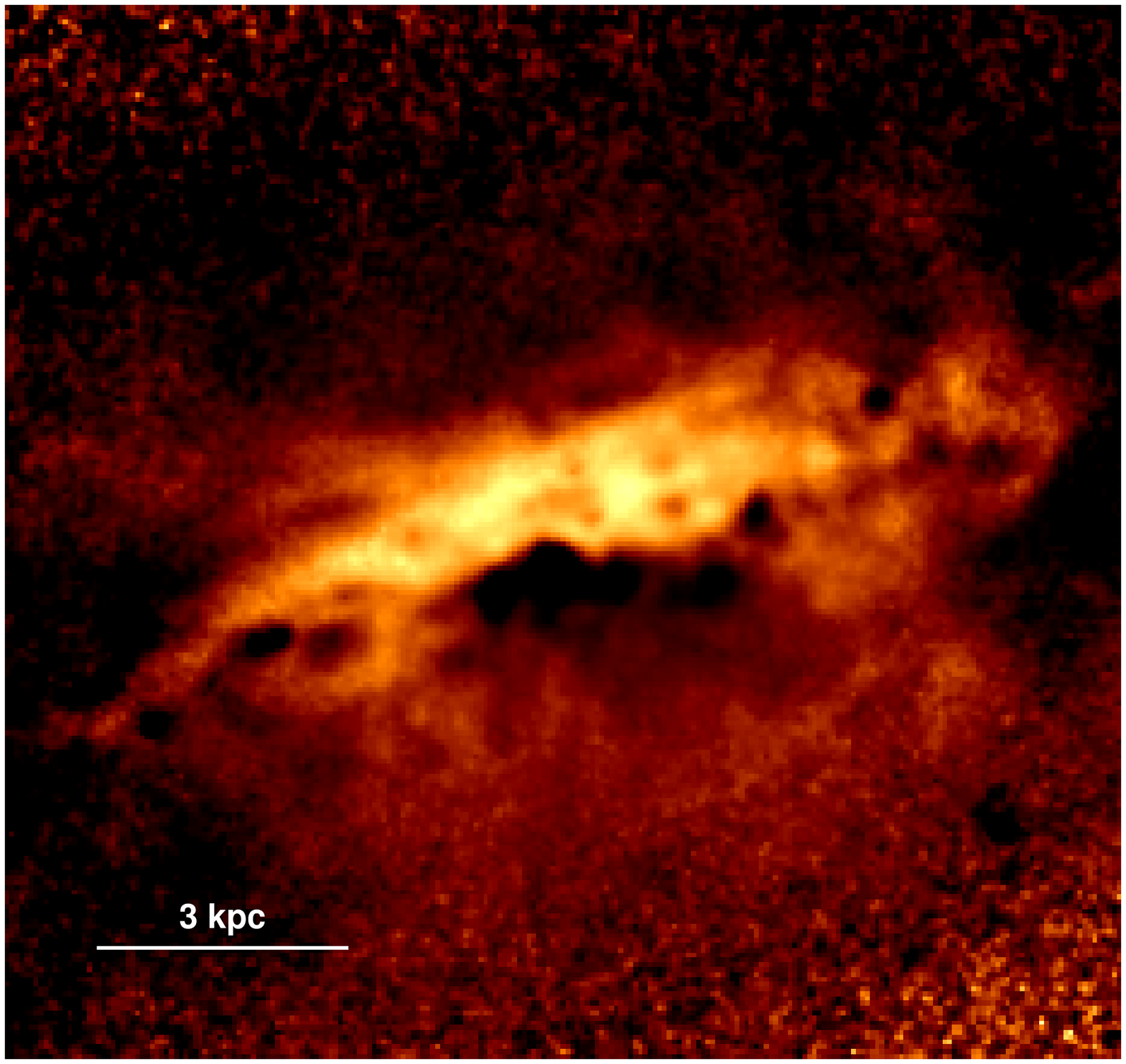}
\caption{ (\textit{left panel}) (B-V) colour-index map derived for NGC 1482. This figure clearly reveals a pair of dust lanes running parallel to the optical major axis of the host galaxy. In addition to the main lanes, filamentary dust structures, originated from the central starburst are also evident.
(\textit{right panel}) Dust extinction map in B band derived after comparing light distribution in the original and modelled image. Brighter shades in both the figures indicate regions associated with dust features.}

\label{fig2}
\end{figure}
\section{Results}
\subsection{Dust properties}
\subsubsection{Dust morphology}
Though this galaxy is known to host a prominent dust lane along its optical major axis (\citep{1991Sci...254..592D}, \citep{2004ApJ...606..829S}, \citep{2005MNRAS.356..998H}), quantitative investigations on the extinction properties of dust contained in this galaxy are not available. In order to investigate extinction properties, accurate knowledge of the extent as well as morphology of dust in the target galaxy is required. For knowing the spatial distribution of dust in NGC 1482 we have generated its ($B-V, B-R, V-R$) colour-index maps by comparing light distribution in the geometrically aligned, seeing matched images in different pass bands (see \citep{2007A&A...461..103P} for details). One of such colour index ($B-V$) image is shown in Figure~\ref{fig2}(left-panel), where brighter shade represent dust occupied regions. This figure clearly reveals a thick, prominent dust lane running along optical major axis of NGC 1482. In addition to the main dust lane, a secondary dust lane, almost parallel to the former, is also evident in this colour-index map. A more careful look at this figure reveals several relatively fainter filamentary dust structures stretched along the minor axis of the NGC 1482 and are analogous to those seen in the \Ha emission Figure~\ref{fig4}(left panel).

\begin{figure}
\center 
\includegraphics[width=50mm,height=50mm]{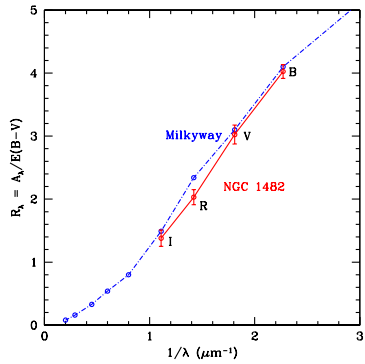}
\caption{%
The average extinction curve derived for the dust occupied region within the NGC 1482 (shown in solid line). For comparison, the Galactic curve is also drawn (dotted line)}
\label{fig3}
\end{figure}

\begin{figure}
\center
\includegraphics[width=45mm,height=40mm]{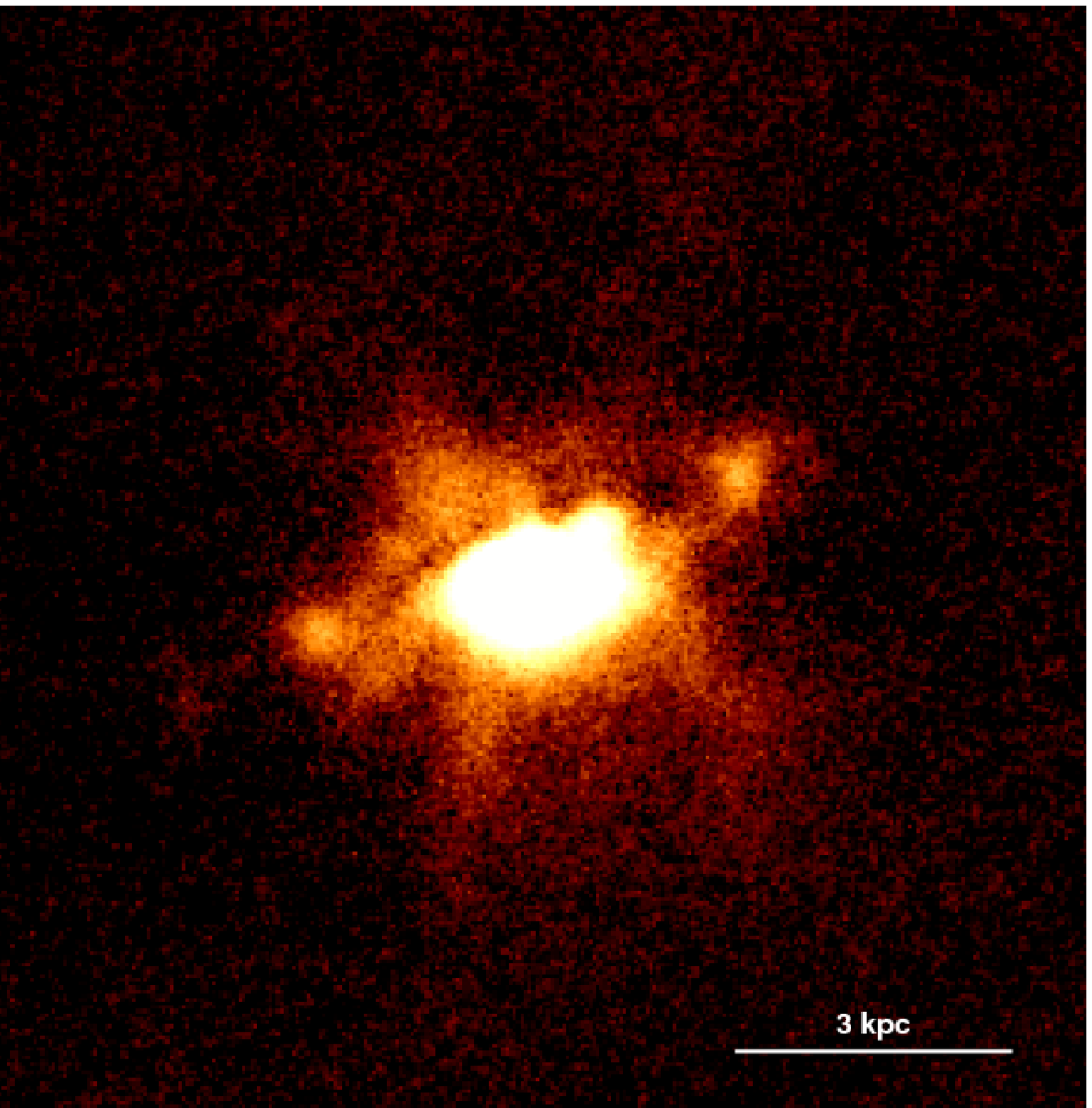}
\includegraphics[width=45mm,height=40mm]{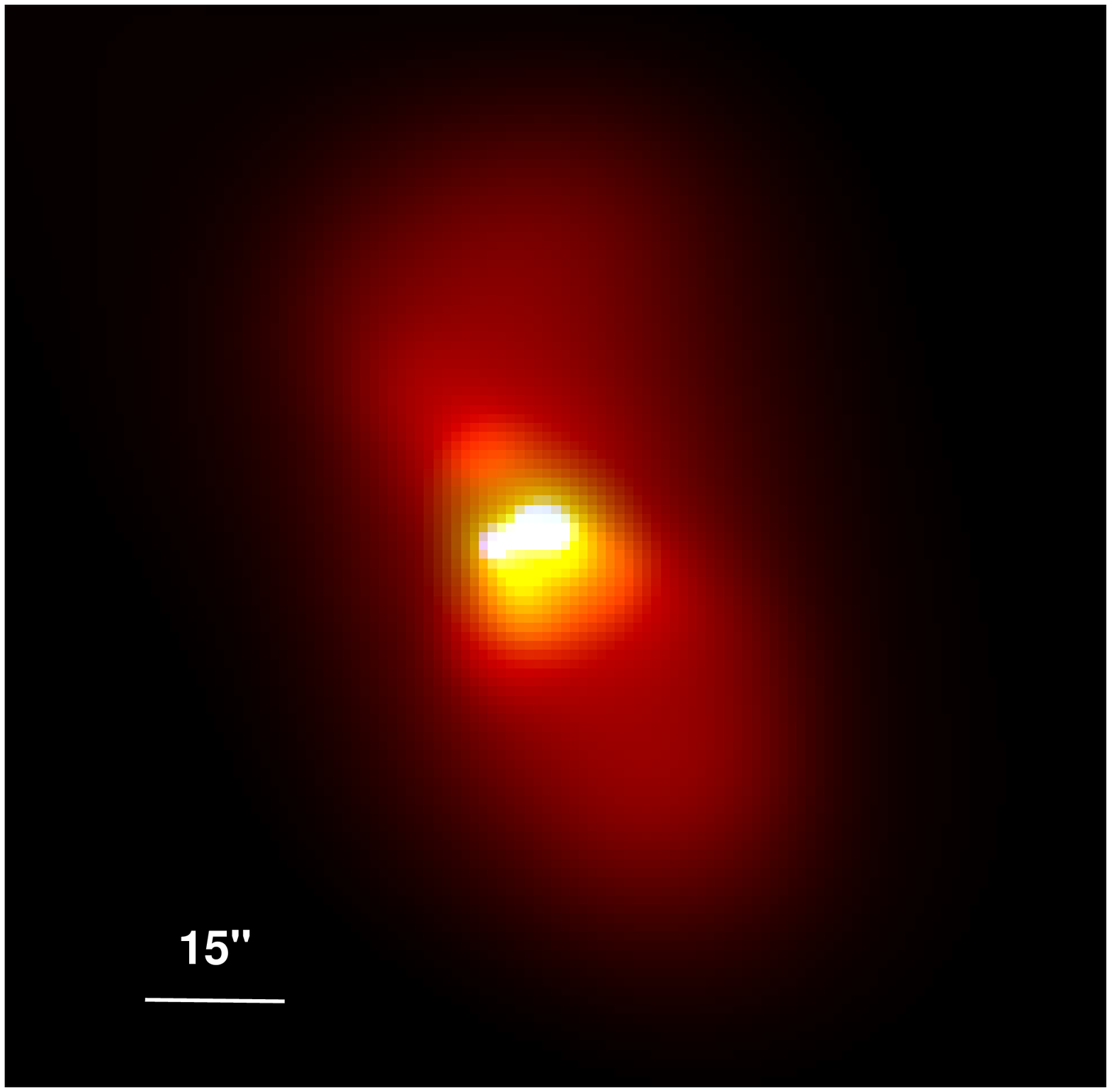}
\caption{\textit{left panel} Continuum subtracted $H_\alpha$ emission-line map derived for NGC 1482. This image clearly exhibit spatial correspondence of the emission-line features with those seen in dust maps.
 {\textit{right panel}} Adaptively smoothed true colour image of NGC 1482. The image is produced by the mosaic of three X-ray band images, namely, the 0.3-1\,keV (red), 1-2\,keV (green) and 2-8\,keV (blue) energy ranges. Notice the extension of diffuse component and the compactness of the nuclear hard component in this figure.}
\label{fig4}
\end{figure}

The process of deriving the wavelength dependent nature of dust extinction (\textit{extinction curve}) in this starburst galaxy NGC 1482 is briefly described here and we suggest the reader to refer \citep{2007A&A...461..103P} for more details. To determine the extinction in different pass bands, one has to compare the actual light distribution in the observed galaxy with that expected in the absence of dust. The method adopted here assumes that elliptical and lenticular galaxies have a smooth and symmetric light distribution with respect to their nuclei. Therefore, it is possible to derive dust free models of  NGC 1482 by fitting ellipses to the light distribution within the image in different passbands. Any deviations seen in the global brightness profile can be recognised by comparing the light distribution in the actual and model images and can be attributed to the obscuration by dust. The quantum of extinction can be measured by computing how much light is missing in the suspected regions relative to the best fitted smooth profile.

We have fitted ellipses to the isophotes of the galaxy images using the \textit{ellipse} fitting program available in the \textbf{stsdas.analysis.isophote} package running within the IRAF image processing software and is based on the procedure outlined by \citep{1987MNRAS.226..747J}. The position angle, ellipticity, and centre of ellipses were free parameters during the fit, as long as the signal reaches the 3$\sigma$ level of the background. Before the model fit were executed, regions occupied by foreground stars were masked and ignored in the further analysis. The model image constructed using the best fit ellipses was then subtracted from the original image and its residual image was generated. Given the existence of dust features in the residual image, we reiterated the ellipse fitting procedure once again, now with the dusty regions masked and discarded from the fit. This resulted in to the dust free model of the original galaxy with smooth underlying stellar light distribution.

\subsubsection{Extinction curve}  
To investigate quantitative properties of the dust extinction in NGC 1482, we constructed its \textit{extinction maps} using the final \textit{dust free} models generated above in the manner, 
\[
A_{\lambda}  = -2.5 \times log\left(\frac{I_{\lambda,obs}}{I_{\lambda,model}}\right)
\]
where A$_\lambda$ is the amount of extinction in a particular pass bands (B, V, R \& I) calculated in magnitude scale, while $I_{\lambda,obs}$ and $I_{\lambda,model}$ are the observed (attenuated) and unextinguished (dust free model) light intensities, respectively, in a given passband. One of such extinction maps derived for NGC 1482 in B band is shown in Figure~\ref{fig2}(right panel), where brighter shades represent regions of higher optical depths associated with the dust extinction. This figure confirms the dust morphology revealed in the ($B-V$) colour-index map including the filamentary features along the minor axis.

We have extracted numerical values of local extinctions in different pass bands ($A_\lambda$) using the extinction maps derived above and then we fitted linear regressions between the local extinction values $A_\lambda$ and the selective extinction ($A_B - A_V$ ) values. The best-fitting slopes of these regressions and their associated uncertainties were subsequently used to derive the $R_\lambda\,\left[=\frac{A_\lambda}{A_B-A_V}\right]$ values and hence the extinction curve.  The average extinction curve derived for the dust occupied regions in the starburst galaxy NGC 1482 is shown in Figure~\ref{fig3} (solid line), along with the Galactic curve (dotted line) for comparison (taken from \citep{1985ApJ...288..618R}). From this figure it is clear that the extinction curve derived for NGC 1482 is very similar to the Galactic curve with the $R_V$ value equal to 3.05 $\pm$ 0.01, little smaller than the Galactic value of 3.1, and is consistent with those reported by \citep{1994A&AS..105..341G}, \citep{2007A&A...461..103P}, \citep{2010MNRAS.409..727F}).  The extinction curve in Figure~\ref{fig3} imply that the grain size distribution within the dusty regions of the starburst galaxy NGC 1482 is almost similar to that of our Galaxy.
\subsubsection{Dust mass estimation}
The total extinction values measured above are used to estimate mass of the dust hosted by this galaxy following the method described by \citep{1994A&AS..105..341G}, \citep{1994MNRAS.271..833G} and \citep{2007A&A...461..103P}. this method assumes that the chemical composition of the dust grains is uniform throughout the galaxy and is similar to that of the canonical grains in the Milky Way. We use the Mathis, \citep{1977ApJ...217..425M} two-component model consisting of spherical silicate and graphite grains with an adequate mixture of sizes, to measure the total dust content in this galaxy. The dust mass was estimated by integrating the dust column density over the dust occupied area (A), yielding: 
\[M_d = A\times\Sigma_d = A\times l_d \times \int_{a_{min}}^{a_{max}} \frac{4}{3}\pi a^3 \rho_d\, n(a) da\] 
where $\Sigma_d$ is the dust column density (g cm$^{-2}$); $\rho_d$ is the specific grain density ($\sim$ 3 g cm$^{-3}$); $l_d$ is the length of the dust column along the line of sight;  $a$ is the grain size; $n(a)=n_0\, a^{-3.5}$; {\bf where n$_0$=A$_i$\,n$_H$, A$_i$ represents the overall abundace of silicate \& graphite} ({\bf in cm$^{2.5}$}) {\bf \& n$_H$ is number density of H nuclei (in cm$^{-3}$) \citep{1984ApJ...285...89D}}; and $a_{min}$ \& $a_{max}$ represent the lower and upper cut-offs of the grain size distribution, respectively. Using the measured value of total extinction in V band ($A_V$), we estimate the dust mass for this galaxy to be equal to 2.7$\times 10^5$ \Msun. As the optical extinction assumes the screening effect of dust, and hence provides a lower limit to the true dust content of the host galaxy. This method is insensitive to the intermix component of dust and is not accounted for in the estimate. The uncertainties involved due to the lower and upper cutoffs in the grain size may also worsen this estimate.

The dust grains exposed to high radiation field from young, massive stars get heated and hence can re-radiate as a black body at longer wavelengths. Therefore, alternatively, the dust content in this galaxy was also be derived using the observed IRAS flux densities ($F_{\nu}$) at 60 $\mu$m and 100 $\mu$m and employing the equation \citep{1989ApJS...70..699Y},
\[ M_{dust} = \frac{4}{3} a \rho_d D^2 \frac{F_\nu}{Q_\nu B_\nu (T_d)} \]
where $D$ luminosity distance of the galaxy in Mpc; $Q_\nu$ and $B_\nu (T_d)$ are the grain emissivity factor and the Planck function for the dust temperature ($T_d$) at frequency $\nu$, respectively \citep{1983QJRAS..24..267H}. The dust grain temperature was calculated using the IRAS flux densities at 60 and 100 $\mu$m and the single-colour equation $T_d = 49\left( \frac{S_{60}}{S_{100}} \right) ^{0.4}$ K \citep{1989ApJS...70..699Y} and is found to be equal to 44 K. This value is higher compared to dust grain  temperature in elliptical galaxies but is consistent with the values reported for starburst galaxies (\citep{1986A&A...154L...8C} and \citep{1999noao.prop....8C}), and imply that dust in starburst galaxies is relatively "hot". The dust mass estimated from the IRAS flux density is $\sim 1.2\times 10^6$ \Msun \, and is higher than the estimate from optical extinction by a factor of 4.4. This discrepancy in the two estimates is due to the fact that out optical method is insensitive to the intermix component of dust, while IRAS can detected them. Therefore, estimate from IRAS flux densities represent true dust content of the target galaxy. Further, this galaxy is also detected at still longer wavelengths (up to 1300$\mu$m) \citep{1995A&A...295..317C}, and hence may enhance this discrepancy if the cold dust hosted by this galaxy is taken in to account.

\subsection{Morphology of "warm" gas}
\Ha emission maps for NGC 1482 were already presented by \citep{2002ApJ...565L..63V}, \citep{2004ApJ...606..829S}. However, to examine association of dust with the ionised gas, we have derived its \Ha emission maps from our narrow band imaging analysis. To derive the pure emission map of NGC 1482, a PSF matched, properly scaled R band image was subtracted as a continuum from the H$_\alpha$ image, and is shown in Figure~\ref{fig4}(left panel). A casual inspection of this figure reveals a close association with the dust in the central region including the filamentary structures seen in the (B-V) colour map. Morphology of the ionised gas derived here is consistent with those reported by \citep{2004ApJ...606..829S}, \citep{2002ApJ...565L..63V}.
\subsection{X-ray gas}
\subsubsection{X-ray emission maps}
\begin{figure} 
\center
\includegraphics[width=80mm,height=55mm]{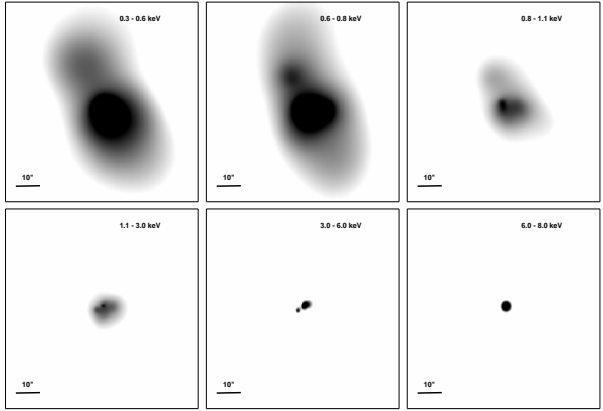}
\caption{%
Diffuse X-ray emission maps in six different energy bands, the energy range is indicated at the top right of each panel. This figure illustrates the evolution of different regions as a function of the energy band. It is apparent that the hard X-ray component is relatively compact and originates from the nuclear region only.}
\label{fig5}
\end{figure}
X-ray emission due to outflows from SNe and stellar winds in this starburst galaxy were already reported in past studies (\citep{2002ApJ...565L..63V}, \citep{2002ApJ...568..689S}. An attempt is made to examine the association of multiphase ISM in this galaxy by comparing multiwavelength observations. For this purpose we derive hot gas emission maps of NGC 1482. Figure~\ref{fig1} shows the 0.3-8\,keV raw image of NGC 1482 at the full resolution (0.\arcsec49/pixel) of the \textit{Chandra} ACIS-S3. From this figure it is clear that the X-ray structure of NGC 1482 is complex and exhibit a bright nuclear region and very extended diffuse emission along its minor axis. Dark patch seen along the direction of dust lane indicate absorption of the soft X-ray photons. Figure~\ref{fig4}(\textit{right panel}) displays a true colour image of NGC 1482 produced by mosaic of three different X-ray bands (red: 0.3-1\,keV, green: 1-2\,keV and blue: 2-8\,keV). These images in different bands were initially adaptively smoothed using the CIAO tool \textit{csmooth} at 3$\sigma$ significance level. From this figure it is clear that the extended emission along the optical minor axis of NGC 1482 is mainly due to the 0.3-1.0\,keV soft component, while the hard X-ray emission (2-8\,keV) is compact and confined to the nuclear region only. The intermediate band X-ray emission (1-2\,keV) are seen around the  hard compact region. 

Figure~\ref{fig5} illustrates the evolution of the nuclear region of NGC 1482 as a function of energy range of X-ray photons in a more clear manner. From this figure it is evident that the hard X-ray emission ($\>3$\,keV) is originating from the the compact nuclear region, while the soft emission with some bright clumps is found to originate from the diffuse component extended along optical minor axis of the host galaxy. As X-rays become harder, this extended component along northeast and southwest direction become fainter at the same time central region becomes brighter. This confirms the origin of the the hard component from the nuclear region. The hardest component in the energy range of 6.0 - 8.0\,keV is coming only from the brightest of the nuclear point source (Reg2 in Figure~\ref{fig7}).
\begin{figure}
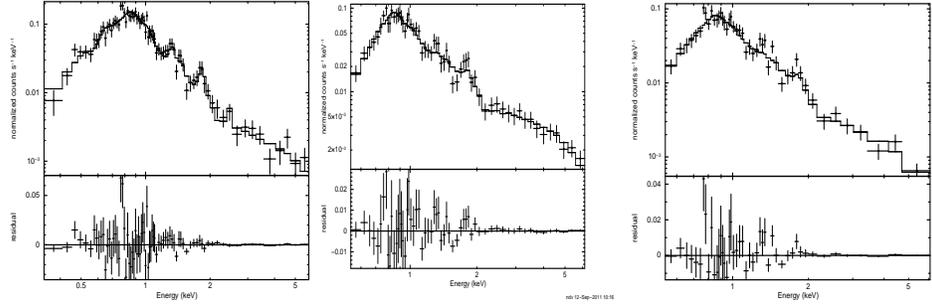
 
\center
\includegraphics[width=40mm,height=40mm]{fig6a.ps}
\includegraphics[width=40mm,height=40mm]{fig6b.ps}
\includegraphics[width=40mm,height=40mm]{fig6c.ps}
\caption{%
(\textit{left panel}) Fitted spectrum of the diffuse X-ray emission extracted from elliptical annulus covering the entire outflow region in NGC 1482. This was fitted with an absorbed vmekal+powerlaw model. (\textit{middle panel}) and (\textit{right panel}) Fitted spectra of the X-ray emission from central 10\arcsec region including both sources and excluding them, respectively. These spectra were fitted with an absorbed \textit{mekal + power law} model. Emission lines due to metals are evident in all the spectra.}
\label{fig6}
\end{figure}
\subsubsection{Spectral analysis}
With the available sub arcsecond resolution of \textit{Chandra} telescope, apart from the global properties, it is also possible to investigate the spatially resolved properties of X-ray emission from this galaxy. This will enable us to constrain on the various processes involved in the X-ray emission and also to disentangle contribution of hard components. For the analysis of events in the 0.3-8.0\,keV range, the XSPEC (v 12.6) spectral-fitting package was used. To ensure applicability of the $\chi^2$ statistic, spectra were binned as to obtain at least 20 counts per fitting bin before background subtraction. To check consistency of the fitting parameters, for the nuclear sources resulting in fewer than 5 degrees of freedom, unbinned spectra were fitted using C-statistic. 
\begin{figure} 
\center
\includegraphics[width=45mm,height=45mm]{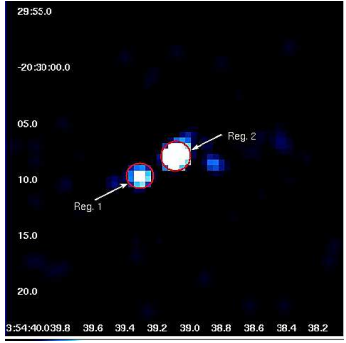}
\caption{%
3-8\,keV X-ray emission map revealing a pair of point-like sources separated by 3.\arcsec 5}
\label{fig7}
\end{figure}

To examine the global properties of hot gas in the target galaxy, we have extracted a point source and background subtracted spectrum of the diffuse emission within the elliptical annulus covering the entire halo part of NGC 1482 and the resulting spectrum is shown in Figure~\ref{fig6}(left panel). For background subtraction we have used properly scaled blank-sky files provided by the CXC. Although there is a strong evidence for spatial variations in temperature (Figure 3b \& 4), we found that a single thermal model adequately represents the target galaxy's flux averaged properties. This spectrum was fitted using the standard $\chi^2$ statistic from $XSPEC$, with an absorbed single temperature component plus power law (wabs$_{Gal}$(wabs(VMEKAL+Power law))). The absorbing column density was considered to be of two components, one due to the ISM within the Milky Way (Galactic column density, $N_H=3.7\times10^{20}$ cm$^{-2}$, \citep{2000AJ....120.2965S}) and the other due to the ISM of NGC 1482. This analysis resulted in to an absorption corrected flux of $4.25^{+0.21}_{-1.44} \times10^{-13}$ erg cm$^{-2}$ s$^{-1}$, which corresponds to the total X-ray luminosity, in the 0.3-8.0\,keV band, of $3.83^{+0.05}_{-1.91} \times 10^{40}$ erg s$^{-1}$. The fitted column density for NGC 1482 was found to be $6.57\pm\,1.10 \times 10^{20}$ cm$^{-2}$, with the thermal component having temperature equal to $0.61\pm0.03$ keV and a fitted abundance of O=1.23$\pm$\,0.25, Ne=0.62$\pm$\,0.32, Mg=1.42$\pm$\,0.39, Si=1.44$\pm$\,0.61, S=3.72$\pm$\,1.93 and Fe=0.41$\pm$\,0.03 all relative to $Z_\odot$, other abundances were frozen at solar values. The power law component had a fitted photon index of $\Gamma = 1.24\pm0.14$. The errors shown here are based on the 90\% confidence level. These estimates are in agreement with those reported by \citep{2004ApJ...606..829S}.

From Figure~\ref{fig5} \&~\ref{fig7} it is obvious that the 3-8\, keV hard X-ray emission is mainly concentrated in the central compact region. Therefore, to study the properties and contribution of nuclear sources we have extracted the 0.3-8\,keV spectrum of the total X-ray emission (resolved plus unresolved) from the inner 10\arcsec circular region and is shown in Figure~\ref{fig6}(middle panel). The spectrum is complex and reveals an absorbed power-law at energies above 3\,keV. This model was fitted using an absorbed, single temperature mekal plus power-law model. The neutral hydrogen column density is high and is found to be $n_H=1.50\pm0.16\times10^{21}$ cm$^{-2}$, and the photon index is $\Gamma=1.17\pm0.07$. The absorption corrected X-ray luminosity is $L_X=4.56^{+0.10}_{-1.46}\times 10^{40}$ erg s$^{-1}$. We have also extracted spectrum of the unresolved emission from within the inner 10\arcsec circle, excluding both the resolved sources, and was fitted with an absorbed, single temperature MEKAL plus power-law model Figure~\ref{fig6}(right panel). This resulted into the luminosity $\sim$ $2.51^{+0.10}_{-1.27}\times 10^{40}$ erg s$^{-1}$, and contribute $\sim$ 45\% of the X-ray emission from this region.

\begin{figure}
\center
\includegraphics[width=60mm,height=60mm]{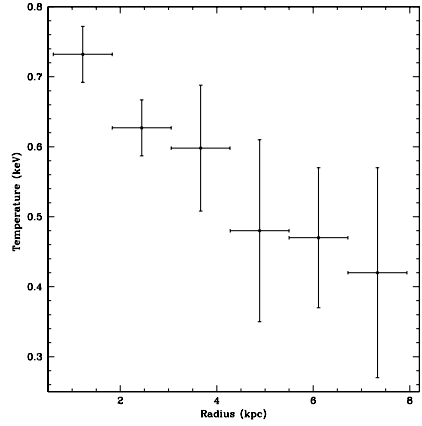}
\caption{Temperature Profile of the X-ray emitting gas extracted from the elliptical annuli centred on NGC 1482. This profile reveals a gradient in the temperature of plasma along the outflow.}
\label{fig8}
\end{figure}

\subsubsection{Point source spectra}
\begin{figure}
\center
\includegraphics[width=40mm,height=40mm]{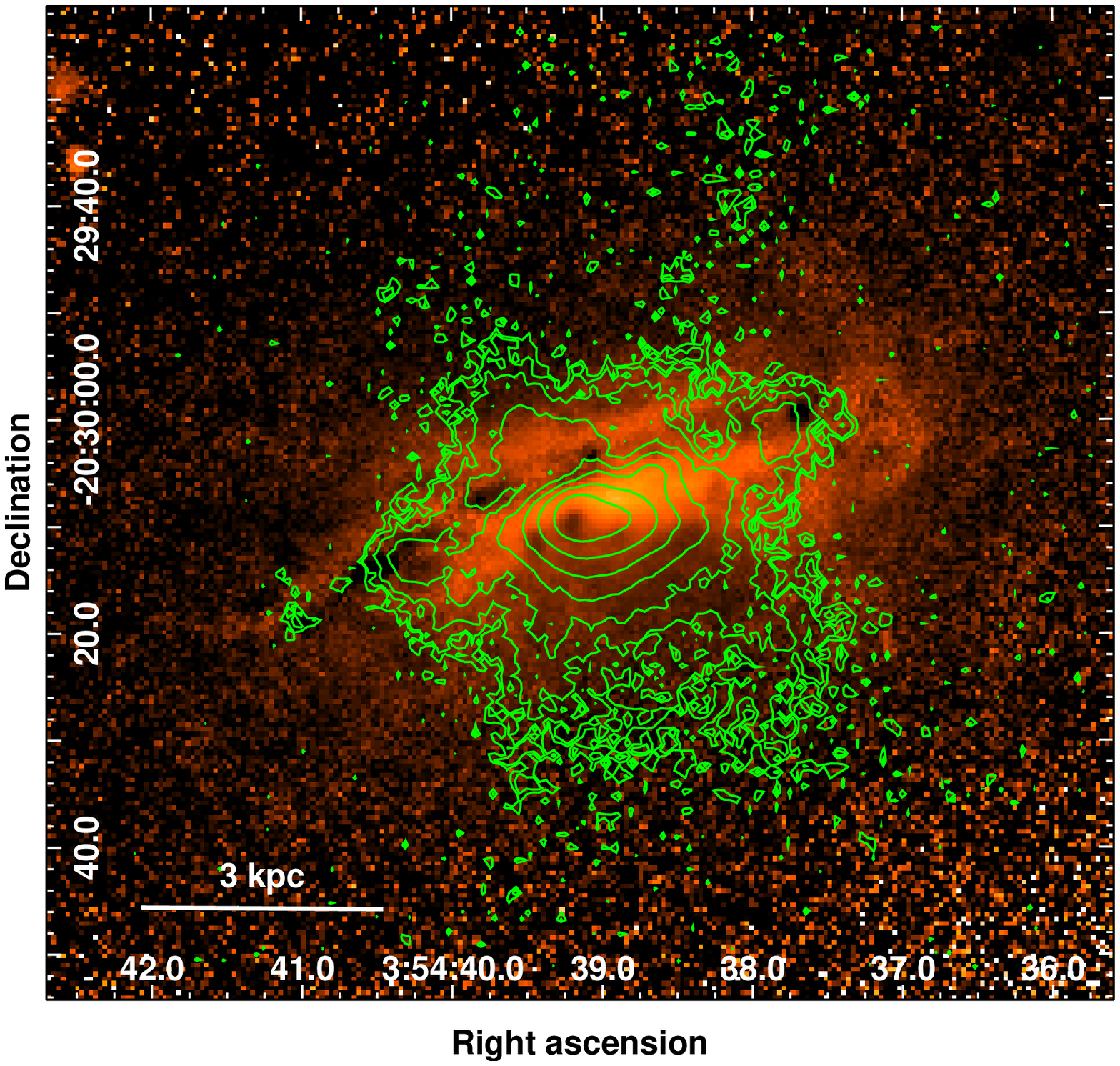}
\includegraphics[width=40mm,height=40mm]{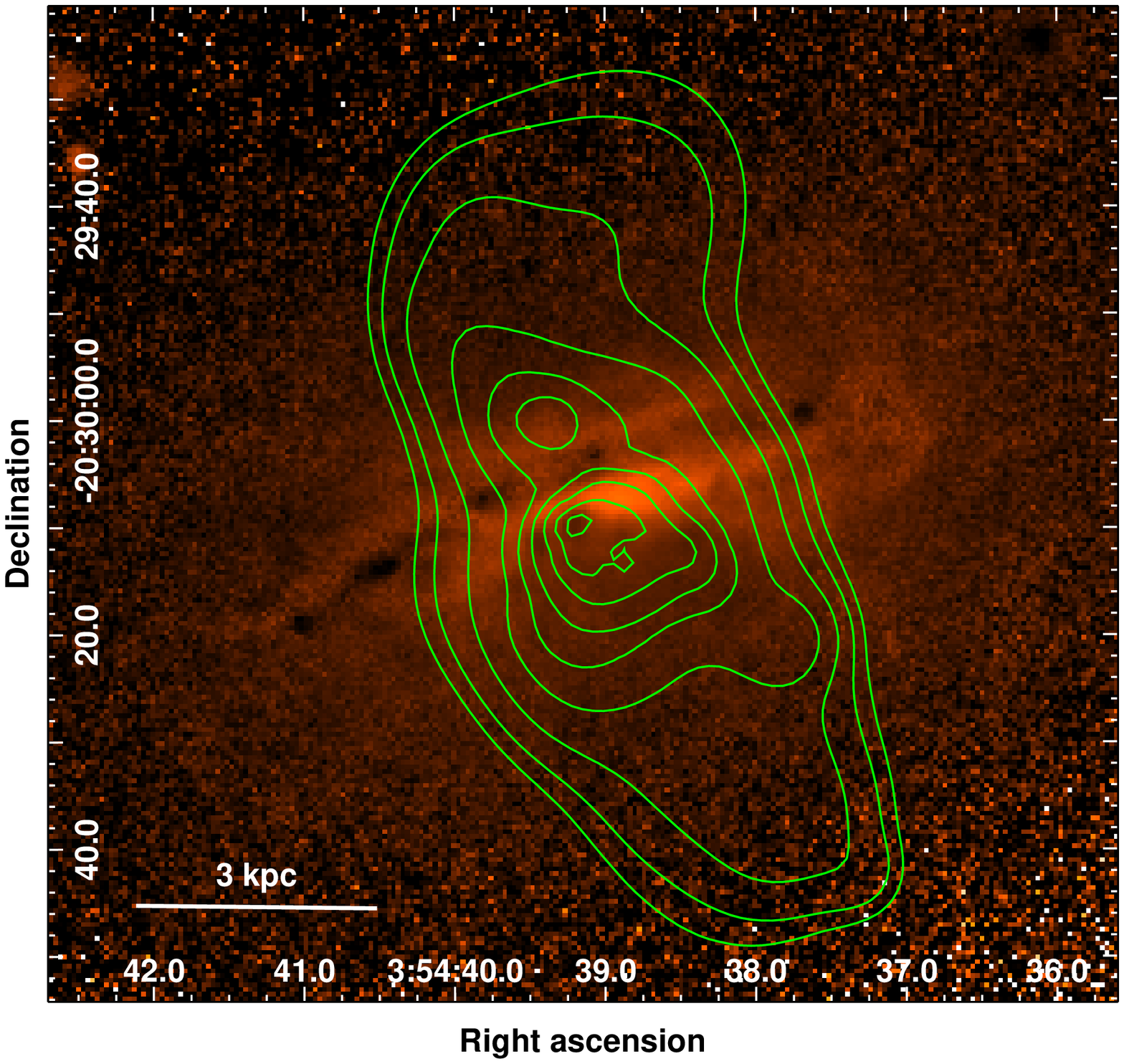}
\includegraphics[width=40mm,height=40mm]{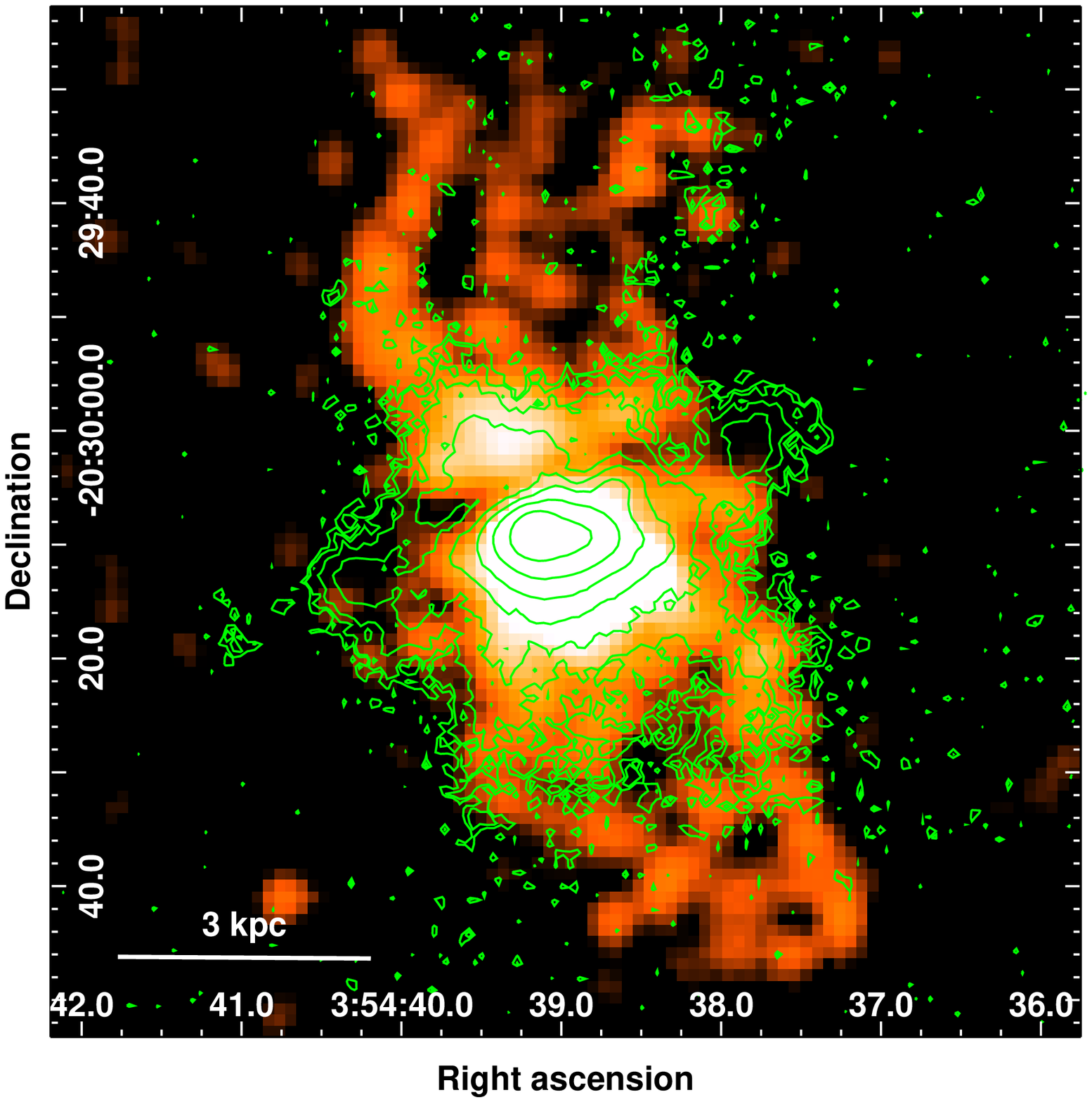}
\caption{(a) (B-V) colour index image derived for NGC 1482, overlaid on which are the \Ha emission line contours. This figure reveals spatial correspondence between the two components. (b) (B-V) colour index image, overlaid on which are the smoothed X-ray contours. (c) X-ray emission map derived after 3$\sigma$ smoothing of X-ray emission, overlaid on it are the \Ha emission line contours. All the three figures exhibit morphological similarities in the three different phases of ISM.}
\label{fig9}
\end{figure}
The true-colour X-ray image Figure~\ref{fig4}(right panel) and the Figure~\ref{fig7} clearly show two bright nuclear sources. We investigated their spectral properties by extracting X-ray photons from individual sources. For this analysis, background spectrum was extracted from the local region. The spectrum of Reg 1 is well fitted (C-stat 10.70/12d.o.f.) by a power law model, with photon index $\Gamma=1.21\pm0.22$. The implied luminosity is $L_X = 2.27\times10^{39}$ erg s$^{-1}$, and is slightly higher than the Eddington luminosity of a 1.5 \Msun accreting compact object.
  
The source and background spectra for Reg 2 were also extracted in the similar way. A simple absorbed power-law model provides a good fit (C-stat 12.51/11d.o.f.), with total column density $n_H=3.54\times10^{22}$ cm$^{-2}$ and power-law photon index $\Gamma=1.67\pm0.11$. The deduced X-ray luminosity is $L_X=9.34\times10^{39}$ erg s$^{-1}$, and is higher than Eddington limit. The spectral fit to the Reg 2 suggest that it lies behind a high column density, and is relatively softer than Reg 1. 

It is known that the amount of visual extinction along a typical line of sight through the ISM is strongly correlated with the total column density of hydrogen \citep{1978ApJ...224..132B}. Hence we estimated the neutral hydrogen column density using the observed values of colour-excess (E(B-V)) over the extinguished areas through the optical imagery using the relation $n_H = 5.8\times 10^{21}\,E(B-V)$ atoms cm$^{-2}$ \citep{1978ApJ...224..132B}. In the present analysis the measured value of optical colour-excess lead to the estimation of $n_H \sim 4\times 10^{21}$ atoms cm$^{-2}$ and is lower than the derived value from spectral analysis. Our estimates of $n_H$ from colour excess matches well with those reported by \citep{2005MNRAS.356..998H} from the radio observations.
\begin{figure}
\center
\includegraphics[width=60mm,height=60mm]{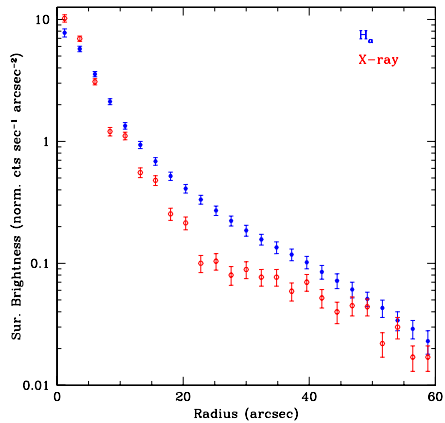}
\caption{%
  Surface brightness profile in X-ray and \Ha region; blue points shows the \Ha and red shows the X-ray}
\label{fig10}
\end{figure}

\begin{figure}
\center
\includegraphics[width=60mm,height=60mm]{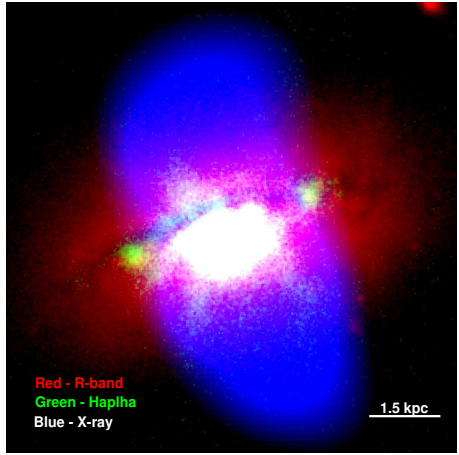}
\caption{%
  Tri-colour image of NGC 1482}
\label{fig11}
\end{figure}

\section{Discussion}
From the X-ray analysis it is apparent that the target galaxy NGC 1482 is embedded in hot ($\sim10^7$\,K) X-ray emitting gas. The dust grains present in such exotic environment are constantly bombarded by high energetic charged particles, that result in erosion or sputtering of the grains. Further, the elevated shocks due to supernovae and stellar winds in the starburst region can also cause for destruction of larger size grains. As a result, dust grains must be undergoing modulation of their sizes and hence one expect relatively smaller grains in starburst galaxies \citep{2003ApJ...598..785N}. This was also confirmed by  \citep{2002AJ....124.3135K}, where they found that crystallinity of the silicate grains is not possible if the effect of supernovae and stellar winds on grain destruction are taken in to account. Given the smaller grains, extinction curve obeyed by them should be gray of flatter than the Milky Way \citep{2000ApJ...533..682C}. However, the average extinction curve derived for this starburst galaxy NGC 1482 using multi-band imaging observations in the optical regime is unexpectedly identical to the Galactic curve (Figure \ref{fig3}), with $R_V$=3.05. This in turn imply that the dust grains responsible for the extinction of starlight in the extreme environment of NGC 1482 are not really different than the canonical grains in the Milky Way. 

For majority of the starburst galaxies, an association between the kpc-scale dust outflows and concentrations of recent star formation within their central disk has been observed (\citep{2004ApJ...606..829S} and the references therein). Additionally, the hot gas pushed out by the massive stars and supernovae in the form of stellar winds is also found to correlate spatially with these two components. An attempt is made to investigate similar association in the target galaxy by examining morphologies of these components. Figure~\ref{fig2}(right panel) shows the (B-V) colour-index map derived for this galaxy, superimposed upon which are the \Ha emission contours. This figure reveals an obvious correspondence between the prominent dust lane as well as the extraplanar dust filamentary complexes and that of \Ha emission. The emission-line gas appears to be considerably more diffuse than that of the extra planar dust and tends to occupy a far greater filling factor above the disk. However, we recognise that our optical extinction method is insensitive to the diffuse component of dust residing above the disk. we also checked the association between the dust and the hot gas distribution, by overlaying diffuse the X-ray emission contours on the (B-V) colour-index map of NGC 1482 (Figure~\ref{fig9}(b)). From this figure it is clear that, though the two components are correlating in the nuclear region, the X-ray emission originating from the SNe and stellar winds have more diffuse and extended distribution. Similar inferences were also derived by past study on this starburst galaxy (\citep{2004ApJ...606..829S}, \citep{2005MNRAS.356..998H}). Figure~\ref{fig11} represents a tri-colour image derived from the contributions of optical R-band image, \Ha emission and X-ray emission, and indicate towards a similar association of the three phases of ISM. 

To examine the spatial correspondence between the X-ray emitting gas and the warm ionised gas in a more quantitative manner, we derive the azimuthally averaged radial surface brightness profile for the X-ray emission and compare it with that from the \Ha emission. For this purpose circular annuli of width 3\arcsec starting from centre of the NGC 1482 out to a radius of 75\arcsec were created and 0.3-2\,keV X-ray photons were extracted from each of the annulus. The azimuthally averaged radial surface brightness profile derived for the diffuse X-ray emission from NGC 1482 is shown in Figure~\ref{fig10} (red data points). We also plot the radial surface brightness profile for the \Ha emission from this galaxy (blue colour points). Comparison of the two profiles imply an obvious similarity between the morphologies of two components, Deviation seen in the X-ray profile in the range between 20-35\arcsec are heavily absorbed by the dust grains seen in the optical imagery of NGC 1482. 

The issue of origin of dust in starburst galaxies is highly controversial. The internal origin assume main contribution from the atmospheres of asymptotic giant branch (AGB) stars. However, AGB stars are too old to account for the presence of dust during the starburst phase of a given galaxy. Therefore, supernovae (SNe) have been recognised as a potential candidate for the injection of dust and gas to the ISM of starburst galaxies, where they can supply $\sim$ 0.1 to 1 \Msun of dust \citep{2007ApJ...662..927D}. The dust injected in to the ISM at the same time is processed  by the forward and reverse shocks in the hot gas swept up by SNe and hence may get destroyed (\citep{2010PhRvD..81h3007N}). With the observed SNe rate in starburst galaxies and considering the two competitive processes of formation and simultaneous destruction of the dust grains, we can estimate total content of dust accumulated by the galaxy over its life time by solving the equation, 
 \begin{center}
 $\frac{\partial M_d(t)} {\partial t}$ = $\frac{\partial M_{d,s}}{\partial t}$ -
 $M_d(t)$ $\it \tau_d^{-1}$ , 
\end{center}
where $\frac{\partial M_d(t)} {\partial t}$ is the dust accumulation rate; ($\frac{\partial M_{d,s}}{\partial t}$) is the rate of dust injection by the SNe and stellar winds; $M_d(t)$ is the available dust at a given time t; and $\tau_d^{-1}$ is the rate of dust grain destruction. Using the measured IR luminosity and the relation given in \citep{2001MNRAS.324..325M}, we estimate the SNe rate for this galaxy to be equal to 0.14 yr$^{-1}$. Assuming that the dust injection has started $\sim 10^8 yr$, and are simultaneously destroyed at $\sim6.0\times10^{-8}\,yr^{-1}$ \citep{1979ApJ...231...77D}, we estimate total dust content of this galaxy $\sim1.1 \times10^5$ {\Msun}. Comparison of this estimate with the true dust content of this galaxy indicate that the injected amount of dust is shorter by an order of magnitude than the actual content. This discrepancy may enhance if we include estimate from the observations at the sub mm wavelength \citep{1995A&A...295..317C}. Thus, the internal supply of  dust is not sufficient to account for the observed dust, and in turn indicate that at least a part of it might have acquired by the host galaxy through a merger like event.

With the given SNe rate, we can examine whether the kinetic energy supplied by the SNe is adequate to drive the observed superwind outflow. Assuming that SNe are supplying energy at the rate $E_{SN}$ ($\sim 10^{51}$ erg), we estimate the total energy injected by the supernova ($L_{kin}$) over the dynamical age of the bubbles ($\sim7.5\times10^6$ yr) using  $L_{kin}=E_{SN} \times R_{SN}$. This gives a total energy supply of $\sim10^{57}$ erg and is consistent with that derived by \citep{2002ApJ...565L..63V} and \citep{2005MNRAS.356..998H}. If we assume that only 10 per cent of it is effectively transmitted to the ISM \citep{1974MNRAS.169..229L}, the net available energy is $\sim10^{56}$ erg. This estimate is higher than the observed energy in the form of X-ray emission ($L_X\sim5.7\times10^{40}$ erg s$^{-1}$). 

Starburst in the central region of NGC 1482 are known to drive several kiloparsec large outflows supplying metal rich gas to the ISM \citep{2004ApJ...606..829S};\citep{2004ApJS..151..193S}. Therefore, to examine the physical properties of this ejecta, we have performed spatially resolved spectral analysis of 0.3-2\,keV X-ray photons extracted from the elliptical annuli centred on NGC 1482. Spectra extracted from each of the annular region were fitted with the \textit{VMEKAL} model. Profile showing variation in temperature of the plasma as a function of distance along the outflow is shown in Figure~\ref{fig8} and indicate gradient in it. This gradient seen in the temperature measurement was also accompanied by the gradient in metallicity value, like those seen in other starburst galaxies. However, these results are not in agreement with those derived by  \citep{2008MNRAS.386.1464R}, where they find increase in the metallicity along the outflow.. The lower value of metallicity in the outer region imply that the diffuse hot gas halo is not yet chemically enriched by the superwinds \citep{2001ApJ...554.1021H}.
\section{Conclusions}
The main conclusions derived from this study are:
\begin{itemize}
\item (B-V) colour-index map derived for the starburst galaxy NGC 1482 confirms two prominent dust lanes running along its optical major axis and are found to extend up to $\sim$ 11\,kpc. In addition to the main lanes, several filamentary structures of dust originating from the central starburst are also evident. Though, the dust is surrounded by exotic environment, the average extinction curve derived for this target galaxy is compatible with the Galactic curve, with $R_V$=3.05, and imply that the dust grains responsible for the optical extinction in the target galaxy are not really different than the canonical grains in the Milky Way. Our estimate of total dust content of NGC 1482 assuming screening effect of dust is $\sim2.7\times10^5$ \Msun, and provide lower limit due to the fact that our method is not sensitive to the intermix component of dust.
\item Adopting present SNe rate for NGC 1482, we quantify total amount of dust supplied by the SNe to the ISM, and was compared with the true dust content of this galaxy from IRAS flux measurements. From this it is observed that the supplied dust is shorter than that actual contained within this starburst galaxy, and hence imply that at least a part of it is acquired by the host galaxy through a merger like event. 
\item Our multiband imaging analysis reveals a qualitative physical correspondence between the morphologies of the \Ha and diffuse X-ray emission was noticed. Similar association is also evident in the kpc-scale dust outflow and its counterparts in the emission line and hot gas.
\item With the known SNe rate for this starburst galaxy, we estimate the amount of kinetic energy supplied by the SNe to the ISM and is found to exceed energy available with the hot gas.
\item Spatially resolved spectral analysis of the hot gas along outflows exhibit a gradient in the temperature. Similar gradient was also noticed in the measured values of metallicity, indicating that the gas in the halo is not yet enriched.
\item High resolution, 2-8\,keV \textit{Chandra} image reveals a pair of point sources in the nuclear region. Absorbed power-law fit to one them gives the hydrogen column density and is higher compared to that estimated from the optical colour excess measurement. The dust mass estimated from the observed X-ray luminosity of this galaxy is in agreement with that estimated from IRAS flux measurements.

\end{itemize}
\section*{Acknowledgments}
We are grateful of the anonymous referee for their careful reading of our manuscript and thoughtful suggestions and comments that have significantly improved the content and clarity of this work. The authors thank the TAC as well as technical Staff of IGO, Pune for their support during observations. This work is supported by ISRO, Bangalore under the RESPOND scheme File No.9/211/2005-II GOI, Dept. of space. This work has made use of X-ray data from the Chandra Data Archive. This publication makes use of the facilities at IUCAA, Pune.


\def\aj{AJ}%
\def\actaa{Acta Astron.}%
\def\araa{ARA\&A}%
\def\apj{ApJ}%
\def\apjl{ApJ}%
\def\apjs{ApJS}%
\def\ao{Appl.~Opt.}%
\def\apss{Ap\&SS}%
\def\aap{A\&A}%
\def\aapr{A\&A~Rev.}%
\def\aaps{A\&AS}%
\def\azh{AZh}%
\def\baas{BAAS}%
\def\bac{Bull. astr. Inst. Czechosl.}%
\def\caa{Chinese Astron. Astrophys.}%
\def\cjaa{Chinese J. Astron. Astrophys.}%
\def\icarus{Icarus}%
\def\jcap{J. Cosmology Astropart. Phys.}%
\def\jrasc{JRASC}%
\def\mnras{MNRAS}%
\def\memras{MmRAS}%
\def\na{New A}%
\def\nar{New A Rev.}%
\def\pasa{PASA}%
\def\pra{Phys.~Rev.~A}%
\def\prb{Phys.~Rev.~B}%
\def\prc{Phys.~Rev.~C}%
\def\prd{Phys.~Rev.~D}%
\def\pre{Phys.~Rev.~E}%
\def\prl{Phys.~Rev.~Lett.}%
\def\pasp{PASP}%
\def\pasj{PASJ}%
\def\qjras{QJRAS}%
\def\rmxaa{Rev. Mexicana Astron. Astrofis.}%
\def\skytel{S\&T}%
\def\solphys{Sol.~Phys.}%
\def\sovast{Soviet~Ast.}%
\def\ssr{Space~Sci.~Rev.}%
\def\zap{ZAp}%
\def\nat{Nature}%
\def\iaucirc{IAU~Circ.}%
\def\aplett{Astrophys.~Lett.}%
\def\apspr{Astrophys.~Space~Phys.~Res.}%
\def\bain{Bull.~Astron.~Inst.~Netherlands}%
\def\fcp{Fund.~Cosmic~Phys.}%
\def\gca{Geochim.~Cosmochim.~Acta}%
\def\grl{Geophys.~Res.~Lett.}%
\def\jcp{J.~Chem.~Phys.}%
\def\jgr{J.~Geophys.~Res.}%
\def\jqsrt{J.~Quant.~Spec.~Radiat.~Transf.}%
\def\memsai{Mem.~Soc.~Astron.~Italiana}%
\def\nphysa{Nucl.~Phys.~A}%
\def\physrep{Phys.~Rep.}%
\def\physscr{Phys.~Scr}%
\def\planss{Planet.~Space~Sci.}%
\def\procspie{Proc.~SPIE}%
\let\astap=\aap
\let\apjlett=\apjl
\let\apjsupp=\apjs
\let\applopt=\ao
\bibliographystyle{mn.bst}
\bibliography{mybib.bib}

\begin{thebibliography}{42}
\expandafter\ifx\csname natexlab\endcsname\relax\def\natexlab#1{#1}\fi

\bibitem[{{Bohlin} {et~al.}(1978){Bohlin}, {Savage}, \&
  {Drake}}]{1978ApJ...224..132B}
{Bohlin}, R.~C., {Savage}, B.~D., \& {Drake}, J.~F., 1978, \apj, 224, 132

\bibitem[{{Calzetti}(2001)}]{2001PASP..113.1449C}
{Calzetti}, D., 2001, \pasp, 113, 1449

\bibitem[{{Calzetti} {et~al.}(2000){Calzetti}, {Armus}, {Bohlin}, {Kinney},
  {Koornneef}, \& {Storchi-Bergmann}}]{2000ApJ...533..682C}
{Calzetti}, D., {Armus}, L., {Bohlin}, R.~C., {Kinney}, A.~L., {Koornneef}, J.,
  \& {Storchi-Bergmann}, T., 2000, \apj, 533, 682

\bibitem[{{Calzetti} {et~al.}(1999){Calzetti}, {Gordon}, \&
  {Clayton}}]{1999noao.prop....8C}
{Calzetti}, D., {Gordon}, K.~D., \& {Clayton}, G.~C., 1999, in NOAO Proposal ID
  \#1999A-0008, pp. 8--+

\bibitem[{{Chini} {et~al.}(1986){Chini}, {Kreysa}, {Mezger}, \&
  {Gemuend}}]{1986A&A...154L...8C}
{Chini}, R., {Kreysa}, E., {Mezger}, P.~G., \& {Gemuend}, H., 1986, \aap, 154,
  L8

\bibitem[{{Chini} {et~al.}(1995){Chini}, {Kruegel}, {Lemke}, \&
  {Ward-Thompson}}]{1995A&A...295..317C}
{Chini}, R., {Kruegel}, E., {Lemke}, R., \& {Ward-Thompson}, D., 1995, \aap,
  295, 317

\bibitem[{{de Vaucouleurs}(1991)}]{1991Sci...254..592D}
{de Vaucouleurs}, G., 1991, Science, 254, 592

\bibitem[{{Draine} \& {Lee}(1984)}]{1984ApJ...285...89D}
{Draine}, B.~T. \& {Lee}, H.~M., 1984, \apj, 285, 89

\bibitem[{{Draine} \& {Salpeter}(1979)}]{1979ApJ...231...77D}
{Draine}, B.~T. \& {Salpeter}, E.~E., 1979, \apj, 231, 77

\bibitem[{{Dwek} {et~al.}(2007){Dwek}, {Galliano}, \&
  {Jones}}]{2007ApJ...662..927D}
{Dwek}, E., {Galliano}, F., \& {Jones}, A.~P., 2007, \apj, 662, 927

\bibitem[{{Finkelman} {et~al.}(2010){Finkelman}, {Brosch}, {Kniazev},
  {V{\"a}is{\"a}nen}, {Buckley}, {O'Donoghue}, {Gulbis}, {Hashimoto},
  {Loaring}, {Romero-Colmenero}, \& {Sefako}}]{2010MNRAS.409..727F}
{Finkelman}, I., {Brosch}, N., {Kniazev}, A.~Y., {et~al.}, 2010, \mnras, 409,
  727

\bibitem[{{Goudfrooij} {et~al.}(1994{\natexlab{a}}){Goudfrooij}, {de Jong},
  {Hansen}, \& {Norgaard-Nielsen}}]{1994MNRAS.271..833G}
{Goudfrooij}, P., {de Jong}, T., {Hansen}, L., \& {Norgaard-Nielsen}, H.~U.,
  1994{\natexlab{a}}, \mnras, 271, 833

\bibitem[{{Goudfrooij} {et~al.}(1994{\natexlab{b}}){Goudfrooij}, {Hansen},
  {Jorgensen}, \& {Norgaard-Nielsen}}]{1994A&AS..105..341G}
{Goudfrooij}, P., {Hansen}, L., {Jorgensen}, H.~E., \& {Norgaard-Nielsen},
  H.~U., 1994{\natexlab{b}}, \aaps, 105, 341

\bibitem[{{Hameed} \& {Devereux}(1999)}]{1999AJ....118..730H}
{Hameed}, S. \& {Devereux}, N., 1999, \aj, 118, 730

\bibitem[{{Hartwell} {et~al.}(2004){Hartwell}, {Stevens}, {Strickland},
  {Heckman}, \& {Summers}}]{2004MNRAS.348..406H}
{Hartwell}, J.~M., {Stevens}, I.~R., {Strickland}, D.~K., {Heckman}, T.~M., \&
  {Summers}, L.~K., 2004, \mnras, 348, 406

\bibitem[{{Heckman} {et~al.}(1990){Heckman}, {Armus}, \&
  {Miley}}]{1990ApJS...74..833H}
{Heckman}, T.~M., {Armus}, L., \& {Miley}, G.~K., 1990, \apjs, 74, 833

\bibitem[{{Heckman} {et~al.}(2001){Heckman}, {Sembach}, {Meurer}, {Strickland},
  {Martin}, {Calzetti}, \& {Leitherer}}]{2001ApJ...554.1021H}
{Heckman}, T.~M., {Sembach}, K.~R., {Meurer}, G.~R., {Strickland}, D.~K.,
  {Martin}, C.~L., {Calzetti}, D., \& {Leitherer}, C., 2001, \apj, 554, 1021

\bibitem[{{Hildebrand}(1983)}]{1983QJRAS..24..267H}
{Hildebrand}, R.~H., 1983, \qjras, 24, 267

\bibitem[{{Hota} \& {Saikia}(2005)}]{2005MNRAS.356..998H}
{Hota}, A. \& {Saikia}, D.~J., 2005, \mnras, 356, 998

\bibitem[{{Jedrzejewski}(1987)}]{1987MNRAS.226..747J}
{Jedrzejewski}, R.~I., 1987, \mnras, 226, 747

\bibitem[{{Jones} {et~al.}(1996){Jones}, {Tielens}, \&
  {Hollenbach}}]{1996ApJ...469..740J}
{Jones}, A.~P., {Tielens}, A.~G.~G.~M., \& {Hollenbach}, D.~J., 1996, \apj,
  469, 740

\bibitem[{{Kewley} {et~al.}(2002){Kewley}, {Geller}, {Jansen}, \&
  {Dopita}}]{2002AJ....124.3135K}
{Kewley}, L.~J., {Geller}, M.~J., {Jansen}, R.~A., \& {Dopita}, M.~A., 2002,
  \aj, 124, 3135

\bibitem[{{Larson}(1974)}]{1974MNRAS.169..229L}
{Larson}, R.~B., 1974, \mnras, 169, 229

\bibitem[{{Mathis} {et~al.}(1977){Mathis}, {Rumpl}, \&
  {Nordsieck}}]{1977ApJ...217..425M}
{Mathis}, J.~S., {Rumpl}, W., \& {Nordsieck}, K.~H., 1977, \apj, 217, 425

\bibitem[{{Mattila} \& {Meikle}(2001)}]{2001MNRAS.324..325M}
{Mattila}, S. \& {Meikle}, W.~P.~S., 2001, \mnras, 324, 325

\bibitem[{{Nozawa} {et~al.}(2010){Nozawa}, {Kohyama}, \&
  {Itoh}}]{2010PhRvD..81h3007N}
{Nozawa}, S., {Kohyama}, Y., \& {Itoh}, N., 2010, \prd, 81, 083007

\bibitem[{{Nozawa} {et~al.}(2003){Nozawa}, {Kozasa}, {Umeda}, {Maeda}, \&
  {Nomoto}}]{2003ApJ...598..785N}
{Nozawa}, T., {Kozasa}, T., {Umeda}, H., {Maeda}, K., \& {Nomoto}, K., 2003,
  \apj, 598, 785

\bibitem[{{Patil} {et~al.}(2009){Patil}, {Pandey}, {Kembhavi}, \&
  {Sahu}}]{2009arXiv0901.1747P}
{Patil}, M.~K., {Pandey}, S.~K., {Kembhavi}, A., \& {Sahu}, D.~K., 2009, ArXiv
  e-prints

\bibitem[{{Patil} {et~al.}(2007){Patil}, {Pandey}, {Sahu}, \&
  {Kembhavi}}]{2007A&A...461..103P}
{Patil}, M.~K., {Pandey}, S.~K., {Sahu}, D.~K., \& {Kembhavi}, A., 2007, \aap,
  461, 103

\bibitem[{{Ranalli} {et~al.}(2008){Ranalli}, {Comastri}, {Origlia}, \&
  {Maiolino}}]{2008MNRAS.386.1464R}
{Ranalli}, P., {Comastri}, A., {Origlia}, L., \& {Maiolino}, R., 2008, \mnras,
  386, 1464

\bibitem[{{Rasmussen} {et~al.}(2004){Rasmussen}, {Stevens}, \&
  {Ponman}}]{2004MNRAS.354..259R}
{Rasmussen}, J., {Stevens}, I.~R., \& {Ponman}, T.~J., 2004, \mnras, 354, 259

\bibitem[{{Rieke} \& {Lebofsky}(1985)}]{1985ApJ...288..618R}
{Rieke}, G.~H. \& {Lebofsky}, M.~J., 1985, \apj, 288, 618

\bibitem[{{Sahu} {et~al.}(1998){Sahu}, {Pandey}, \&
  {Kembhavi}}]{1998A&A...333..803S}
{Sahu}, D.~K., {Pandey}, S.~K., \& {Kembhavi}, A., 1998, \aap, 333, 803

\bibitem[{{Soifer} {et~al.}(1989){Soifer}, {Boehmer}, {Neugebauer}, \&
  {Sanders}}]{1989AJ.....98..766S}
{Soifer}, B.~T., {Boehmer}, L., {Neugebauer}, G., \& {Sanders}, D.~B., 1989,
  \aj, 98, 766

\bibitem[{{Strickland} {et~al.}(2004{\natexlab{a}}){Strickland}, {Heckman},
  {Colbert}, {Hoopes}, \& {Weaver}}]{2004ApJS..151..193S}
{Strickland}, D.~K., {Heckman}, T.~M., {Colbert}, E.~J.~M., {Hoopes}, C.~G., \&
  {Weaver}, K.~A., 2004{\natexlab{a}}, \apjs, 151, 193

\bibitem[{{Strickland} {et~al.}(2004{\natexlab{b}}){Strickland}, {Heckman},
  {Colbert}, {Hoopes}, \& {Weaver}}]{2004ApJ...606..829S}
---, 2004{\natexlab{b}}, \apj, 606, 829

\bibitem[{{Strickland} {et~al.}(2000){Strickland}, {Heckman}, {Weaver}, \&
  {Dahlem}}]{2000AJ....120.2965S}
{Strickland}, D.~K., {Heckman}, T.~M., {Weaver}, K.~A., \& {Dahlem}, M., 2000,
  \aj, 120, 2965

\bibitem[{{Strickland} {et~al.}(2002){Strickland}, {Heckman}, {Weaver},
  {Hoopes}, \& {Dahlem}}]{2002ApJ...568..689S}
{Strickland}, D.~K., {Heckman}, T.~M., {Weaver}, K.~A., {Hoopes}, C.~G., \&
  {Dahlem}, M., 2002, \apj, 568, 689

\bibitem[{{Thornley} {et~al.}(2000){Thornley}, {Schreiber}, {Lutz}, {Genzel},
  {Spoon}, {Kunze}, \& {Sternberg}}]{2000ApJ...539..641T}
{Thornley}, M.~D., {Schreiber}, N.~M.~F., {Lutz}, D., {Genzel}, R., {Spoon},
  H.~W.~W., {Kunze}, D., \& {Sternberg}, A., 2000, \apj, 539, 641

\bibitem[{{Tsuru} {et~al.}(2007){Tsuru}, {Ozawa}, {Hyodo}, {Matsumoto},
  {Koyama}, {Awaki}, {Fujimoto}, {Griffiths}, {Kilbourne}, {Matsushita},
  {Mitsuda}, {Ptak}, {Ranalli}, \& {Yamasaki}}]{2007PASJ...59S.269T}
{Tsuru}, T.~G., {Ozawa}, M., {Hyodo}, Y., {et~al.}, 2007, \pasj, 59, 269

\bibitem[{{Veilleux} \& {Rupke}(2002)}]{2002ApJ...565L..63V}
{Veilleux}, S. \& {Rupke}, D.~S., 2002, \apjl, 565, L63

\bibitem[{{Young} {et~al.}(1989){Young}, {Xie}, {Kenney}, \&
  {Rice}}]{1989ApJS...70..699Y}
{Young}, J.~S., {Xie}, S., {Kenney}, J.~D.~P., \& {Rice}, W.~L., 1989, \apjs,
  70, 699

\end{thebibliography}
\end{document}